\input{aipcheck}

\documentclass[
    ,draft       
  ]
  {aipproc}

\layoutstyle{6x9}

\begin{document}

\title{Diffraction Dissociation - 50 Years later}


\author{Sebastian N. White}{
  address={{Brookhaven National Laboratory, \\ 
        Upton, N.Y. 11973, USA}}
}

\begin{abstract}
The field of Diffraction Dissociation, which is the subject of this workshop, began 50 years ago with the analysis of deuteron stripping in low energy collisions with nuclei. We return to the subject in a modern context- deuteron dissociation in $\sqrt{s_{NN}}= 200$ GeV d-Au collisions recorded during the 2003 RHIC run in the PHENIX experiment. At RHIC energy, d$\rightarrow$n+p proceeds predominantly ($90\%$) through Electromagnetic Dissociation and the remaining fraction via the hadronic shadowing described by Glauber. Since the dissociation cross section has a small theoretical error we adopt this process to normalize other cross sections measured in RHIC.
\end{abstract}

\maketitle


\section{Introduction}
	When deuteron beams were first accelerated to 190 MeV and collided with internal targets in the Berkeley cyclotron, experiments found a very collimated forward beam of neutrons which were identified with the process of absorptive stripping originally proposed by Serber\cite{Serber}. Glauber\cite{Glauber} then showed that deuteron breakup could also proceed via a process he called "free dissociation", which has no classical analog. Absorption of part of the deuteron occurs even when neither nucleon strikes the target nucleus (treated as a black disc) and this absorption mixes unbound states of the proton and neutron which can then appear in the outgoing beam. The calculated
cross section for this process is large (60$\%$ of the absorptive stripping cross section). 

	At RHIC collision energy  ($\sqrt{s_{NN}}= 200$ GeV) a second mechanism for free dissociation of the deuteron becomes dominant. The intense Coulomb field of the target nucleus appears to the incident deuteron as a beam of photons whose flux can be calculated by the Equivalent Photon Approximation originally due to Fermi\cite{fermi,white_erice}. Since the spectrum extends well above the deuteron photodissociation energy ($E_{\gamma}=2.23$ MeV) this becomes the dominant process for free dissociation.  A recent calculation\cite{vogt} for RHIC yields 1.38 ($\pm 5\%$) barn of which 0.14 barn is due to the original nuclear dissociation process calculated by Glauber.
	
	In 2 companion papers in the early 60's Good and Walker\cite{Good} observed that both Coulomb dissociation and nuclear dissociation should have analogs in diffractive excitation of elementary particles in high energy collisions with nuclei. Diffractive processes play a significant role in both Coulomb interactions (in Ultraperipheral Collisions) and hadronic collisions of Heavy Ions. This should be contrasted with e-p (HERA) and $\bar{p}p$ (Tevatron) where either one or the other process is studied. A unique aspect of the Heavy Ion program in ATLAS is that HERA and Tevatron measurements can be extended to higher energies and nuclear targets\cite{hot_topics,ramona}.	
	\section{d-Au Cross Sections}
In addition to the dissociation cross section the total d-Au inelastic cross section is of interest for the RHIC program. The inelastic cross section is sampled in the experiments ("min-bias trigger") for use as a 
luminosity monitor. Once the min-bias cross section is known those of other processes recorded during the same luminosity interval can be calculated in the usual way.
	There are 2 approaches to this cross section normalization. In the first it is derived from known, elementary, NN inelastic cross sections using a Glauber model with a Woods-Saxon distribution parametrization of the p,n distributions in the nuclei. Calculations done for d-Au at RHIC energy range from 2.26$\pm0.1$ barn \cite{Nardi} to 1.93 barn\cite{boris}. It is difficult to assign an overall error to this calculation since one may find discussions in the literature of whether or not to include the diffractive part of NN cross sections and whether n and p should have the same matter distributions in the Au nucleus, etc.
	
		The second approach, which we adopt here, is to directly determine the min-bias trigger cross section by comparison to the reliably calculated \cite{vogt} deuteron dissociation process which was also measured in the PHENIX 2003 data.
		
\section{Instrumentation}
	The four RHIC experiments have mid-rapidity spectrometers with different characteristics but all share identical Zero Degree Calorimeters (ZDC's) located at $\pm 18$ m. The ZDC's cover $\pm 5$ cm (in x and y) about the forward beam direction and have an energy resolution of $\sigma_E/E<21\%$
	for 100 GeV neutrons within x,y$\leq4.5$ cm\cite{Instrum,MCD}. Essentially, all non-interaction ("spectator") neutrons are detected in the ZDC's, while charged particles are mostly swept out of the ZDC region by strong (~16$T\cdot m$) accelerator dipoles at z$=\pm 11$m.
	
		The same dipoles sweep spectator protons from d-dissociation beyond the outgoing beam trajectory (since they have twice the deuteron charge-to-mass ratio) and in PHENIX they are detected in a proton calorimeter("fCal")\cite{pCal}.
		
		The PHENIX experiment used two additional hodoscopes (BBC's)\cite{MCD}, located at z$=\pm1.5$m and covering $3.0\leq|\eta_{BBC}|\leq3.9$, as its main min-bias trigger. Events with one or more charged particles hitting both the +z and -z BBC fired this trigger. Determining $\sigma_{BBC}$ is equivalent to determining the luminosity of the PHENIX data. d-Au events occuring well within the z$=\pm1.5$ m interval between the BBC's fire this trigger with an efficiency of $88\pm4\%$ \cite{eBBC} but the efficiency falls off for $|z_{event}|\geq40$ cm. For this reason we will determine $\sigma_{BBC}$ using only events within this interval. A correction is then applied for the fraction of all RHIC events within this interval. The actual distribution of events along z within the data can be measured using time-of-flight measurements between the ZDC's for events with a ZDC coincidence (with single event resolution of $\sigma_Z\sim 2$ cm).

\begin{figure}
 \includegraphics[height=0.4\textheight]{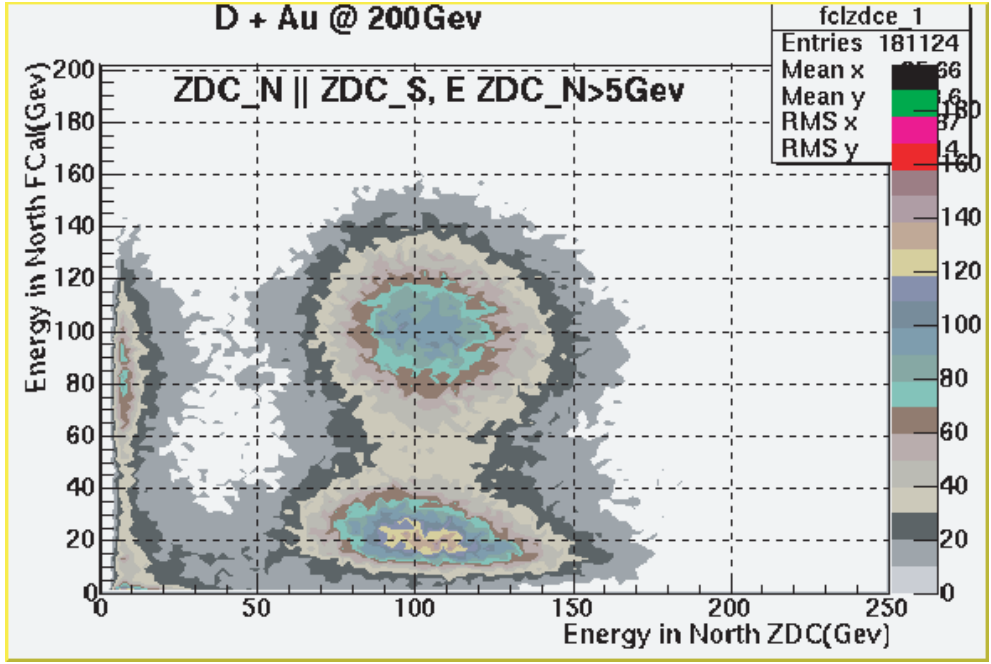}
  \caption{Energy deposition in proton calorimeter vs. ZDC (neutron calorimeter) for events with some 
  activity in the ZDC in the deuteron beam direction. This sample includes absorptive stripping as well as 
  d$\rightarrow$ n+p.}
\end{figure}
\section{Data Analysis}

	Typical event rates were several kHz ($\cal{L}$$\sim1-4\times10^{28}cm^{-2}s^{-1})$ for all processes considered here. Therefore this analysis is based on a representative data sample with 
\begin{equation}
N^{trig}_{BBC}=230k 
\end{equation}
\begin{equation}
N^{trig}(ZDC_{Au}\,"or"\,ZDC_d)= 460k 
\end{equation}
events, where subscripts Au and d represent $E_{ZDC}>10$ GeV in the Au or d direction. The second trigger is sensitive to d-dissociation, characterized by a 100 GeV neutron in $ZDC_d$ and a 100 GeV proton in fCal with no activity at mid-rapidity.
	
	Additional data samples were recorded with one of the RHIC beams intentionally displaced by up to 1 mm to measure the fraction of triggers due to d-Au collisions (as opposed to beam gas background). The largest background was <3$\%$ and quoted rates have been corrected for the measured background.
	
	The BBC rate, corrected for accelerator interaction distribution is
	\begin{equation}
N^{corrected}_{BBC}=228634\, \rm{events}\, (\pm0.5\%)
\end{equation}

\begin{figure}
 \includegraphics[height=0.4\textheight]{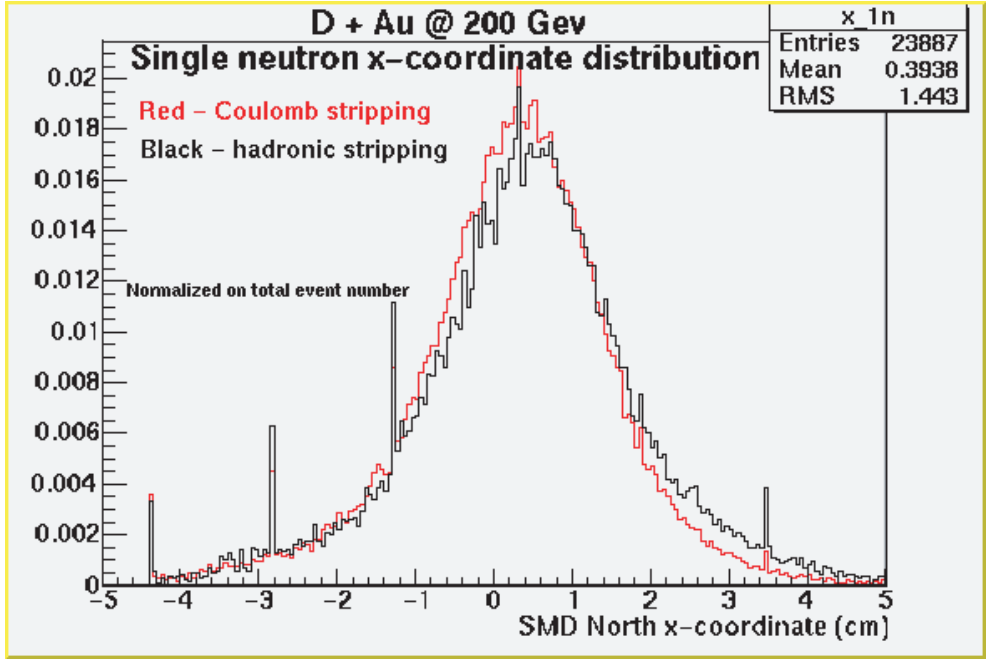}
  \caption{Neutron impact distribution for both absorptive stripping and dissociation events.}
\end{figure}

\section{d-dissociation Analysis}

	As stated above, d-dissociation events have a clear signature in the PHENIX experiment. This is illustrated in Figs 1 where we display the forward (in the deuteron direction) neutron vs. proton calorimeter energy for ZDC-trigger events.
	
		The neutron impact parameter distribution (measured by the PHENIX Shower maximum detector) is displayed in Fig. 2 both for events with the d-dissociation signature ("Coulomb") and BBC trigger ("hadronic"). The latter correspond exactly to Serber's absorptive stripping proccess and, as noted in ref.\cite{boris}, the neutron angular distribution should have an interesting correlation with event centrality. This will be discussed in a future note.
		
		In any case one can see that neutrons have a small angular divergence and consequently there is only a small correction for ZDC acceptance. Instead, the dominant correction is for absorptive stripping events which feed into the dissociation sample. In order to extract the dissociation event yield we used an iterative procedure, fitting the sum of $ZDC_d$+fCal total energy to the sum of 100 +200 GeV lineshapes and correcting for calculated efficiency as successive cuts on activity in other detectors were applied. The first 2 iterations yield 
			\begin{equation}
N(d\rightarrow n+p)=157149\, \rm{and} \,156951
\end{equation}
events, so the procedure is clearly stable.

	Our final result is :
		\begin{equation}
\sigma_{BBC}=N(BBC)/N(d\rightarrow n+p)\times\sigma(d\rightarrow n+p)=228634/158761\times1.38 (\pm0.5\%)
\end{equation}
	\begin{equation}
=1.99\, (\pm1.6\%\pm5.0\%)\, \rm{barn}.
\end{equation}

	This is the quantity needed for luminosity normalization.
	
	In order to compare with Glauber calculations in the literature we then correct for the BBC detector efficiency given above:
	
\begin{equation}
\sigma(inelastic_{d-Au})=\sigma_{corrected}(BBC)=1.99/0.88 
\end{equation}
\begin{equation}
=2.26 (\pm1.6\%\pm5.0\%\pm4.5\%)\, \rm{barn}
\end{equation}
where the last 2 errors reflect the theoretical error on $\sigma(d\rightarrow n+p)$ and the BBC inefficiency uncertainty.

	A similar analysis yields the cross section for the selection $ZDC_{Au}(E>10 GeV)$, also used as a min-bias trigger:
\begin{equation}
\sigma(ZDC_{Au})=2.06(\pm1.7\%\pm5.0\%)\, \rm{barn}.
\end{equation}

\begin{theacknowledgments}
 I would like to thank the organizers of this workshop and particularly Dr. Khoze for a very stimulating and enjoyable meeting. I would like to thank my collaborator in this work, A. Denisov and also R. Glauber for helpful discussions. This work was supported in part under
DOE Contract number DE-AC02-98CH10886.\end{theacknowledgments}



\begin{thebibliography}{9}
\bibitem{Serber} R.~Serber, Phys. Rev. \emph{72}(1947) 1008.
\bibitem{Glauber}
R.~J. Glauber,  \emph{Deuteron Stripping Processes at High Energy},Phys. Rev. \emph{99}(1955) 1515.
\bibitem{Instrum} C.~Adler et al., Nucl. Instr. And Meth. \emph{A470}  (2001) 488.
\bibitem{MCD} M.~Chiu et al. Phys. Rev. Lett. \emph{89} (02) 012302 and nucl-ex/0109018.
\bibitem{vogt} S.~Klein and R. ~Vogt, Phys. Rev. C 68 (2003) 017902. and nucl-ex/0303013.
\bibitem{boris} B. Kopeliovich, Phys. Rev. C 68 (2003) 044906 and nucl-th/0306044.
\bibitem{Good} M.~L.~Good and W.~D.~Walker, Phys.Rev.\emph{120} (1960) pp.1855-1856 and 1857-1858.
\bibitem{white_erice} S.~White "Applications of the Equivalent Photon Approximation to Heavy Ion Collisions" in "Electromagnetic Probes of Fundamental Physics" W. Marciano and S. White, eds. World Scientific (2003).
\bibitem{fermi} E.~Fermi (translation by M. Gallinaro and S. White) "On the Theory of Interactions between Atoms and Electrically Charged Particles", ibid and hep-th/0205086.
\bibitem{hot_topics} G.~Bauer et al. "Hot Topics in Ultraperipheral Heavy Ion Collisions", ibid.
\bibitem{ramona} M.~Strikman, R.V~ogt and S. White, manuscript in preparation. To be published in CERN yellow report on UltraPeripheral Collision physics and R. Vogt ,hep-ph/0407298.
\bibitem{Nardi} D.~Kharzeev, E.~Levin and M.~Nardi, hep-ph/0212316.
\bibitem{eBBC} S. S. Adler et al., (PHENIX Collaboration), Phys. Rev.
Lett. 91, 072303 (2003).
\bibitem{pCal} The fCal calorimeters were re-used from BNL experiment E-864. See  T.A. Armstrong et al., Nucl. Instr. and Meth. A 406 (1998) 227.


\end{thebibliography}
\end{document}